\documentstyle[12pt,fullpage,subeqn]{article}
\begin{document}
\begin{flushright}
TIFR/TH/98-42 \\
SU-4240-687 \\
\end{flushright}
\bigskip\bigskip
\begin{center}
{\Large{\bf MAXIMALLY REALISTIC CAUSAL QUANTUM MECHANICS}}$^{\dagger, \star}$
\\[1cm]
{\large S.M. ROY$^a$ and VIRENDRA SINGH$^a$} 

\vskip 1cm

{\sl Department of Theoretical Physics,} \\ 
{\sl Tata Institute of Fundamental Research,} \\
{\sl Homi Bhabha Road, Mumbai 400 005, INDIA.}
\end{center}

\vskip 5ex

\centerline{\bf ABSTRACT}

\vskip 3ex

We present a causal Hamiltonian quantum mechanics in $2n$-dimensional
phase space which is more realistic than de Broglie-Bohm
mechanics. The positive definite phase space density reproduces as 
marginals the correct quantum probability densities of $n+1$ different
complete commuting sets of observables e.g. positions, momenta and
$n-1$ other sets. 

\vskip 6ex

\noindent PACS: 03.65.Bz

\vskip 2ex

\vfill

\hrule width 3in

\medskip

\noindent $^\dagger$ A preliminary version of this work was presented
by one of us (SMR) at the symposium ``Causality and Locality in Modern
Physics and Astronomy, open questions and possible solutions : A
Symposium to honour Jean-Pierre Vigier'', Aug. 25-29, 1997, York
University, Canada.\\

\noindent $^\star$ Part of the work was done while the authors were
visiting Department of Physics, Syracuse University, Syracuse, N.Y.\\

\noindent$^a$e-mail: shasanka@theory.tifr.res.in , \ \ \ \
vsingh@theory.tifr.res.in 

\newpage

\noindent 1. \underbar{Quantum Contextuality}.  The most important
difference between quantum mechanics and a classical stochastic theory
is that quantum probabilities are inherently and irreducibly context
(i.e., experimental arrangement) dependent.  For any complete
commuting set (CCS) of observables $A$, the quantum state
$|\psi\rangle$ specifies the probability of observing the eigenvalues
$\alpha$ as $|\langle \alpha|\psi \rangle |^2$, \underbar{if $A$ were to be
observed}.  If $B$ is another CCS with eigenvalues $\beta$, but $[A,B]
\neq 0$, the analogous probabilities $|\langle \beta|\psi\rangle|^2$,
\underbar{if $B$ were to be measured} refer to a different context or
experimental situation.  Each context corresponds to the experimental
arrangement to measure one CCS of observables.  Due to this
\underbar{inherent} context dependence quantum mechanics does not
specify joint probabilities of noncommuting observables.  Moreover,
the context dependence is irreducible, i.e., quantum mechanics cannot
be embedded in a classical context independent stochastic theory.
This is the lesson from decades of work, e.g. Gleason's theorem$^1$,
Kochen-Specker theorem$^2$, Bell's theorem$^3$ (where the
contextuality corresponds to violation of ``local realism'') and
Martin-Roy theorem$^4$ (which is a direct phase space proof relevant
to the present work).  The contextuality theorems circumscribe the
extent to which dynamical variables in quantum mechanics can be
ascribed simultaneous `Reality' independent of observations. 
\bigskip

\noindent 2. \underbar{De Broglie-Bohm}.  The De Broglie-Bohm (dBB)
causal quantum mechanics$^5$ has shaped a paradigm of causal quantum
mechanics in which the Position Observable occupies a favoured status
of a ``beable''$^6$ with values independent of context or observation,
whereas other observables may have context dependent values.  The
state of the individual system is characterized by $\{|\psi(t)\rangle,
\vec x(t)\}$ where $|\psi(t)\rangle$ and $\vec x(t)$ are the state
vector and the configuration space coordinates, whereas an ensemble of
these states corresponds to $|\psi(t)\rangle$.  For a many particle
system with the quantum Hamiltonian
\[
H = -\sum_i {\hbar^2 \over 2m_i} \nabla^2_i + U(\vec x),
\]
the individual $\vec x_i(t)$ move according to 
\[
(\vec p_i)_{\rm dBB} = m_i {d\vec x_i \over dt} = \vec\nabla_i S(\vec
x(t),t), 
\]
where $\psi = R \exp(i S/\hbar)$, with $R$ and $S$ being real and
$\vec x$ denoting $(\vec x_1, \vec x_2, \cdots )$.  This
means that the phase space dynamics is determined by the causal
Hamiltonian 
\[
H_{\rm dBB} (\vec x,\vec p,t) = \sum_i {{\vec p}^2_i \over 2m_i} +
U(\vec x) - \sum_i {\hbar^2 \over 2m_i R} \vec\nabla^2_i R
\]
and corresponds to the phase space density
\[
\rho_{\rm dBB} (\vec x,\vec p,t) = |\psi(\vec x,t)|^2 \delta(\vec p -
\vec p_{\rm dBB} (\vec x,t)).
\]
(We shall set $\hbar = 1$ henceforth).  Integration over momentum
shows that the ensemble position density agrees with $|\psi(\vec
x,t)|^2$.  However, as pointed out by Takabayasi$^7$, integration over
position does not yield the quantum momentum density $|\tilde\psi(\vec
p,t)|^2$, where $\tilde\psi$ is the Fourier transform of $\psi$.  This
disagreement exhibits the context dependence of momentum in dBB
thoery: the preexisting momentum probability density given by the dBB
theory is assumed to be converted into the correct quantum density by
means of a measurement interaction appropriate to the context of a
momentum measurement.  On the other hand, position measurements simply
reveal the existing position. Thus the position measurement
interaction does not play the same role of altering the existing
probability distribution.

\bigskip

\noindent 3. \underbar{Motivations For A Causal Quantum Mechanics More
Realistic than De Broglie-Bohm} \hfill\break \underbar{Theory}.  The
asymmetrical treatment of position and momentum constitutes the
breaking of a fundamental symmetry of quantum theory which has
sometimes been considered as a defect of the dBB theory (Holland,
Ref. 5, p. 21).  We recently constructed a causal quantum
mechanics$^{8, 9}$ in one dimension, in which Takabayasi's objection as
well as the asymmetric treatment of position and momentum are removed.
Without invoking the measurement interaction, the new causal theory
reproduces both position and momentum probability distributions of
usual quantum theory, and is therefore more realistic than the dBB
theory.  

Can we formulate a notion of a maximally realistic causal mechanics
(for spinless particles with a configuration space of $n$ dimensions)
which respects quantum contextuality theorems?  A mechanics which
yields Hamiltonian evolution of phase space variables with a positive
definite phase space density will be called a `Causal Hamiltonian
Mechanics' or a `Causal Mechanics' in brief.  A `Causal Mechanics'
which simultaneously reproduces the quantum probability densities of
the maximum number of different (mutually noncommuting) CCS of
obserables as marginals of the same phase space density will be called
a `Maximally Realistic Causal Quantum Mechanics'.  The definition is
nontrivial because the contextuality theorems do not allow probability
distributions of all possible CCS to be simultaneously reproduced. 

What constraints can we impose selfconsistently on the phase space
probability density $\rho(\vec x,\vec p,t)$ of an ensemble of phase
space points of a causal theory?  Motivated by the success in one
dimension$^{8, 9}$ we may require that the quantum position and momentum
probability densities are reproduced as `marginals', i.e.,
\begin{equation}
\int \rho(\vec x,\vec p,t) d\vec p = |\psi(\vec x,t)|^2,
\end{equation}
\begin{equation}
\int \rho(\vec x,\vec p,t) d\vec x = |\tilde \psi(\vec p,t)|^2,
\end{equation}
where $\tilde \psi$ denotes the Fourier transform of the wave function
$\psi$.  The probability interpretation necessitates the positivity
condition, 
\begin{equation}
\rho(\vec x,\vec p,t) \geq 0,
\end{equation}
which rules out many phase space distribution functions such as the
Wigner function.$^{10}$  Moreover, positive distribution functions
obtained by smoothing the Wigner function$^{11}$ do not in general
reproduce the correct marginals.  Neverthless, as emphasized by Cohen
and Zaparovanny$^{12}$, the uncertainty principle does not preclude
the existence of a phase space density obeying conditions (1) -- (3).
A simple example is
\[
\rho_0 (\vec x,\vec p,t) = |\psi(\vec x,t)|^2 |\tilde\psi(\vec
p,t)|^2. 
\]
For a causal theory a further condition is necessary if the phase
space density is to arise from an underlying Hamiltonian dynamics,
viz. the ``Liouville condition'',
\begin{equation}
{\partial\rho \over \partial t} (\vec x,\vec p,t) + \sum^n_{i=1}
\left(\dot x_i {\partial\rho \over \partial x_i} + \dot p_i {\partial\rho
\over \partial p_i}\right) = 0,
\end{equation}
where a dot denotes time-derivative.  As is well known, this condition
is an immediate consequence of the phase space continuity eqn.
$$
{\partial \rho(\vec x,\vec p,t) \over \partial t} + \sum^n_{i=1}
\left\{{\partial \over \partial x_i} (\dot x_i \rho) + {\partial \over
\partial p_i} (\dot p_i \rho)\right\} = 0
\eqno (4a)
$$
and the existence of a causal Hamiltonian $H_c(\vec x,\vec p,t)$ such
that 
$$
\partial H_c/\partial p_i = \dot x_i, \ \partial H_c/\partial x_i =
-\dot p_i. 
\eqno (4b)
$$

The achievement of de Broglie and Bohm$^5$ was to construct a causal
mechanics obeying (1), (3) and (4). The mechanics we
constructed$^{8, 9}$ for $n=1$ is more realistic because it obeys
Eq. (2) in addition.  The new mechanics has the phase space density
\begin{equation}
\rho (x,p,t) = |\psi(x,t)|^2 |\psi(p,t)|^2 \delta
\left(\int^p_{-\infty} dp' |\psi(p',t)|^2 - \int^{\epsilon
x}_{-\infty} dx' |\psi(\epsilon x',t)|^2\right)
\end{equation}
where $\epsilon = \pm 1$, and we have omitted the tilda denoting
Fourier transform (i.e., $\psi(p,t)$ actually stands for $\tilde
\psi(p,t) = \langle p|\psi(t)\rangle$).  Eqs. (1), (2) and (3) are
obviously satisfied; further, it also has a $c$-number causal
Hamiltonian of the form$^{8, 9}$
\[
H_c = {1 \over 2m} (p - A(x,t))^2 + V(x,t)
\]
with two quantum potentials $A$ and $V$ (instead of just one in the de
Broglie-Bohm theory) which depend on the wave function $\psi$.  Hence
the Liouville condition (4) is also obeyed.

We shall see that in higher dimensions, we can and should be even more
ambitious. 

\bigskip

\noindent 4. \underbar{Maximally Realistic Causal Quantum Mechanics}.
The purpose of the present work is to construct a new mechanics which
we tentatively call ``maximally realistic causal quantum mechanics''
in 2n-dimensional phase space. In this theory a single phase space
probability density reproduces the quantum probability densities of
$n+1$ different CCS of observables in spite of the fact that no two
sets are mutually commuting. The pleasant surprise is that not only
the quantum probability densities of position and momentum, but also
those of $n-1$ other CCS of observables can be simultaneously
realized. The choice of the $n+1$ CCS whose probabilities are
simultaneously realized is not unique; the appropriate choice can
depend on the context. Different contexts have the same wave function
but different phase space probability densities; thus the
wave function is not a complete description of these
probabilities. 

The phase space quantum contextuality theorem of Martin and Roy$^4$
proves that it is impossible to realize quantum probability densities
for all possible choices of the CCS of observables as marginals of one
positive definite phase space density. We conjecture that the
simultaneous realization of quantum probability densities of more than
$n+1$ different CCS is impossible and hence that the causal theory
presented here is maximally realistic.

Let the state of the individual system be characterized by $\{|\psi
(t) > , \vec x (t)\}$ or $\{|\psi (t) > , \vec p (t) \}$ since any $n$
independent phase space coordinates are now on equal footing. Due to
the freedom of canonical transformations, we may assume without loss
of generality that the CCS $(X_1, \cdots X_n)$ is among the $n+1$ CCS
whose quantum probability densities are reproduced in the new causal
theory. We assume (without any fundamental justification) a one-to-one
relation between coordinates and momenta, as this played a crucial
role in our construction of a causal hamiltonian in one
dimension$^{8, 9}$. The phase space density must then be of the general
form, 
\begin{equation}
\rho (\vec x, \vec p, t) = |\psi (\vec x, t)|^2 \prod^n_{j=1} \delta
(p_j - \hat p_j (\vec x, t)) 
\end{equation}
which returns the correct marginal $|\psi (\vec x, t)|^2$ on
integration over the momenta. We shall now show that the functions
$\hat{\vec p} (\vec x, t)$ can be chosen so as to reproduce the quantum
probability densities of a `chain' of $n+1$ different CCS, e.g.
\begin{equation}
(X_1, X_2, \cdots , X_n), (P_1, X_2, \cdots , X_n), (P_1, P_2, X_3,
\cdots , X_n), \cdots , (P_1, P_2, \cdots , P_n) ,
\end{equation}
where each CCS in the chain is obtained from the preceding one by
replacing one phase space variable by its canonical conjugate. In the
one dimensional case the rquirement of one-to-one relation between
$x$ and $\hat p (x,t)$ yields two discrete solutions $\hat p (x, t)$
(corresponding to $\epsilon = \pm 1$ in Eq. (5)) which are non-decreasing
and non-increasing functions of $x$ respectively. Analogously, in the
n-dim case, there is a 2-fold ambiguity in determining the phase space
variables of each CCS in the chain (7) in terms of the preceding CCS,
and hence $2^n$ discrete solutions $\hat{\vec p} (\vec x, t)$. These
solutions can be read off from the $\delta$-functions in the $2^n$
phase space densities, each of which reproduces the desired $n+1$
quantum probability densities as marginals:
\begin{equation}
\rho (\vec x, \vec p, t) = \prod^n_{i=0} |\psi (\Omega_i, t)|^2 \
\prod^n_{j=1} \delta (A_j) .
\end{equation}
Here, each $\Omega_i$ denotes phase space variables corresponding to
one CCS :
\begin{eqnarray}
\Omega_0 &=& (x_1, x_2, \cdots , x_n) , \nonumber \\ 
\Omega_i &=& (p_1, p_2, \cdots , p_i , x_{i+1}, \cdots , x_n) , \ \
{\rm for} \ \ 1 \leq i \leq n-1 , \nonumber \\
\Omega_n &=& (p_1, p_2, \cdots , p_n) , \nonumber
\end{eqnarray}
and the $\psi(\Omega_i,t)$ denote appropriate Fourier transforms of
$\psi(\Omega_o,t)$.  Each $\delta (A_j)$ serves to determine
$\Omega_j$ in terms of $\Omega_{j-1}$ ,
\begin{eqnarray}
A_1 &=& \int^{p_1}_{-\infty} |\psi (p'_1, x_2, \cdots , x_n, t)|^2
dp'_1 - \int^{\epsilon_1x_1}_{-\infty} |\psi (\epsilon_1 x'_1, x_2,
\cdots , x_n, t)|^2 dx'_1 , \nonumber \\ 
A_j &=& \int^{p_j}_{-\infty} |\psi (p_1, \cdots , p_{j-1}, p'_j,
x_{j+1} , \cdots , x_n, t)|^2 dp'_j \nonumber \\
&& - \int^{\epsilon_j x_j}_{-\infty} |\psi (p_1, \cdots , p_{j-1},
\epsilon_j x'_j, x_{j+1} , \cdots , x_n, t)|^2 dx'_j , \ \ {\rm for} \
\ 1 < j < n , \nonumber \\
A_n &=& \int^{p_n}_{-\infty} |\psi (p_1, \cdots , p_{n-1}, p'_n,t)|^2
dp'_n \nonumber \\ 
&& - \int^{\epsilon_n x_n}_{-\infty} |\psi (p_1, \cdots , p_{n-1},
\epsilon_n x'_n, t)|^2 dx'_n , \nonumber 
\end{eqnarray}
with
\[
\epsilon_i = \pm 1 , \ \ \ {\rm for} \ \ 1 \leq i \leq n .
\]
Since there are $2^n$ possible choices of the $\epsilon_1, \cdots ,
\epsilon_n$ we have here $2^n$ phase space densities. Direct
integration over $n$ variables, (using the $n$ $\delta$-functions),
yields 
\[
\int \rho (\vec x, \vec p, t) d \bar\Omega_i = |\psi (\Omega_i, t)|^2 
\]
which are the correct marginals. Here $\bar\Omega_i$ denotes the
$n$-tuple of phase space variables complementary to $\Omega_i$, i.e.,
$\bar\Omega_i = (x_1, x_2, \cdots , x_i, p_{i+1}, \cdots , p_n)$, with
$\bar\Omega_0 = (p_1, \cdots , p_n)$ and $\bar\Omega_n = (x_1, x_2,
\cdots , x_n)$. The condition $A_1 = 0$ determines $p_1$ in terms of
$x_1, x_2, \cdots , x_n$ and $t$, i.e., $\hat p_1 (x_1, \cdots , x_n,
t)$; $A_2 = 0$ determines $p_2$ in terms of $p_1$ and $x_2, \cdots ,
x_n, t$ and hence $\hat p_2 (x_1, \cdots x_n, t)$ after substituting
$p_1 = \hat p_1$, and so on. Hence all the momenta are determined via
the $\delta (A_j)$ in terms of $x_1, \cdots , x_n, t$, and the phase
space density (8) can be rewritten in the form (6). (The
$\delta$-functions $\delta (A_j)$ can of course be used equally well
to determine the coordinates in terms of momenta. E.g. $A_n = 0$
yields $x_n$ in terms of $p_1, \cdots , p_n, t$; $A_{n-1} = 0$ yields
$x_{n-1}$ in terms of $p_1, \cdots , p_{n-1}, x_n, t$ and hence in
terms of $p_1, \cdots , p_n, t$ after substituting for $x_n$, and so
on).

The phase space density (8) corresponds to the choice (7) of the
$n+1$ CCS. The form (6) is however more general since it results for
any choice of the chain of $n+1$ CCS which includes $(X_1, X_2,
\cdots, X_n)$. E.g. for $n=2$, all the three chains $\{(X_1, X_2),
(P_1, X_2), (P_1, P_2)\}, \{(X_1, P_2),$ $(X_1, X_2), (P_1, X_2)\}$, and
$\{(X_1, X_2), (X_1, P_2), (P_1, P_2)\}$ will lead to Eq. (6), of
course with different functions $\hat{\vec p} (\vec x, t)$.

\bigskip

\noindent\underbar{Consistency Condition on Velocities due to
Schr\"odinger Eqn.} \ The density $\rho (\vec x, \vec p, t)$ of the
ensemble of system points depends on the $n+1$ marginals $|\psi
(\Omega_i, t)|^2$. Hence the velocities of the individual system
points will also be constrained by the time dependent Schr\"odinger
Eqn. We work out these constraints explicitly when the Schr\"odinger
Eqn. is of the form :
\[
i\hbar \partial \psi /\partial t = \left( \sum_i {P^2_i \over 2m_i} +
U (\vec x) \right) \psi 
\]
in terms of the chosen coordinates and momenta. Starting from the
general form (6) of the phase space density and taking a partial 
derivative w.r.t. $t$ with $\vec x$ and $\vec p$ fixed we obtain, 
\begin{eqnarray}
{\partial \over \partial t} \rho (\vec x, \vec p, t) &=&
\left({\partial \over \partial t} |\psi (\vec x, t)|^2 \right)
\prod^n_{j=1} \delta (p_j - \hat p_j (\vec x, t)) \nonumber \\
&& - \sum^n_{i=1} {\partial \over \partial p_i} \left[{\partial \hat
p_i (\vec x, t) \over \partial t} \rho (\vec x, \vec p, t)\right] .
\end{eqnarray}
The time dependent Schr\"odinger Eqn. yields the probability current
conservation eqn. 
\begin{equation}
{\partial \over \partial t} |\psi (\vec x, t)|^2 + \sum^n_{i=1}
{\partial \over \partial x_i} j_i (\vec x, t) = 0 , 
\end{equation}
where
\begin{equation}
j_i (\vec x, t) = {\rm Re} \left[ \psi^\star (\vec x, t) {-i \over m_i} \
{\partial \over \partial x_i} \ \psi (\vec x, t) \right] .
\end{equation}
Further, the conservation of the number of system points in phase
space yields the phase space continuity eqn. (4a) for $\partial\rho /
\partial t$, with $\dot x_i = \dot x_i (\vec x, \vec p, t), \ \dot p_i
= \dot p_i (\vec x, \vec p, t)$. Substituting Eqs. (4a) and (10)
into Eq. (9) we obtain, 
\begin{eqnarray}
&& \sum^n_{i=1} \left[ {\partial \over \partial x_i} \left( v_i |\psi
(\vec x, t)|^2 - j_i (\vec x, t)\right)\right] \prod^n_{j=1} \delta
(p_j - \hat p_j (\vec x, t)) \nonumber \\ [2mm]
&&~~~~~ + {\partial \over \partial p_i} \left\{ \left({dp_i \over dt} -
{d\hat p_i (\vec x, t) \over dt}\right) |\psi (\vec x, t)|^2
\prod^n_{j=1} \delta (p_j - \hat p_j (\vec x, t))\right\} = 0 ,
\end{eqnarray}
where $\vec v$ deotes the system point velocities. Thus 
\begin{equation}
\vec v (\vec x, t) = \dot{\vec x} (\vec x, \vec p, t) \bigg|_{\vec p =
\hat{\vec p} (\vec x, t)} ,
\end{equation}
\[
{\partial \over \partial x_i} \left(v_i |\psi (\vec x, t)|^2\right) =
\left( {\partial \over \partial x_i} + \sum^n_{k=1} {\partial \hat p k
\over \partial x_i} \ {\partial \over \partial p_k}\right) \left(\dot
x_i |\psi (\vec x, t)|^2\right) \bigg|_{\vec p = \hat{\vec p}} ,
\]
and
\[
{d\hat p_i (\vec x, t)\over dt} = {\partial \hat p_i (\vec x, t) \over
\partial t} + \sum^n_{k=1} {\partial \hat p_i (\vec x, t) \over
\partial x_k} \dot x_k (\vec x, \vec p, t) \bigg|_{\vec p = \hat{\vec
p}} 
\]
Integrating Eq. (12) after multiplying by $dp_1 \cdots dp_n$ we obtain
the constraint on velocities, 
\begin{equation}
\sum^n_{i=1} {\partial \over \partial x_i} \left(v_i |\psi (\vec x, t)|^2 -
j_i (\vec x, t)\right) = 0 .
\end{equation}
Similarly, multiplying Eq. (12) by $p_k dp_1 \cdots dp_n$ and
integrating, we obtain
\[
\left({dp_k \over dt} - {d\hat p_k (\vec x, t) \over dt}\right)
\bigg|_{\vec p = \hat{\vec p} (\vec x, t)} = 0 ,
\]
which is identically satisfied. 

For $n=1$, Eq. (14) implies that the dBB velocity is the unique
solution if we wish to avoid singularities of the velocity at nodes of
the wave function.  For $n > 1$, Eq. (14) can be solved for the
velocities to yield,
\begin{equation}
(v_i - v_{i,B}) |\psi (\vec x, t)|^2 = \sum_\ell {\partial
W_{i\ell} (\vec x, t) \over \partial x_\ell} ,
\end{equation}
where
\subequations
\begin{equation}
W_{i\ell} = - W_{\ell i} , 
\end{equation}
and $v_{i,B}$ denotes the $dBB$ velocity
\begin{equation}
v_{i,B} = j_i (\vec x, t) / |\psi (\vec x, t) |^2
\end{equation}
\endsubequations
The velocities given by (15) differ from the dBB velocities due to the
term involving the antisymmetric tensor $W$.  Eq. (15) was derived
directly from Eq. (10) by Deotto and Ghirardi and by Holland in their
search for atternatives to dBB trajectories$^{13}$.  Our derivation
shows that the Schr\"odinger Eqn. places no other constraints, for
example on $\hat p_j(\vec x,t)$.  We shall now see that in
general $W$ has to be non-zero in order that a causal
Hamiltonian exist. 
\bigskip

\noindent\underbar{Partial Differential Equations For Velocities From
Existence of the Causal Hamiltonian}.  In addition to the constraint
(14) due to the Schr\"odinger Eqn., the velocities must also obey
partial differential eqns. which follow from the requirement of
existence of a $c$-number causal Hamiltonian $H_c(\vec x,\vec p,t)$.
If $\dot x_i$ and $\dot p_i$ are derived via Hamilton's equations,
\begin{equation}
v_k(\vec x,t) = \dot x_k\big|_{\vec p = \hat{\vec p}} =
\left({\partial H_c (\vec x,\vec p,t) \over \partial
p_k}\right)\Big|_{\vec p = \hat{\vec p}},
\end{equation}
\begin{equation}
{d\hat p_i (\vec x,t) \over dt} = {dp_i \over dt}\Big|_{\vec p =
\hat{\vec p}} = - {\partial H_c (\vec x,\vec p,t) \over
\partial x_i}\Big|_{\vec p = \hat{\vec p}}.
\end{equation}
Defining
\begin{equation}
\hat H_c (\vec x,t) = H_c (\vec x,\vec p,t)\big|_{\vec p = \hat{\vec
p}(\vec x,t)},
\end{equation}
and substituting Eqs. (17) and (18), we have,
\begin{eqnarray}
{\partial \hat H_c (\vec x,t) \over \partial x_i} &=& \left({\partial
H_c \over \partial x_i} + {\partial \hat p_k \over \partial x_i}
{\partial H_c \over \partial p_k}\right)\Big|_{\vec p = \hat{\vec p}}
\nonumber \\[2mm] &=& - {d \hat p_i (\vec x,t) \over dt} + {\partial
\hat p_k \over \partial x_i} v_k (\vec x,t).
\end{eqnarray}
In order that a function $\hat H_c (\vec x,t)$ obeying the partial
differential eqns. (20) exist, the integrability conditons 
\begin{equation}
{\partial^2 \hat H_c \over \partial x_i \partial x_j} = {\partial^2
\hat H_c \over \partial x_j \partial x_i}
\end{equation}
must hold.  Substituting (20) into (21) we obtain the $n(n-1)/2$
conditions on the velocities, $(1 \leq i < j \leq n)$,
\begin{equation}
{\partial \over \partial x_i} (f_{kj} v_k) - {\partial \over \partial
x_j} (f_{ki} v_k) + {\partial \over \partial t} f_{ij} (\vec x,t) = 0,
\end{equation}
where
\begin{equation}
f_{ij} (\vec x,t) = {\partial \over \partial x_i} \hat p_j (\vec x,t)
- {\partial \over \partial x_j} \hat p_i (\vec x,t).
\end{equation}
Note that these partial differential eqns. for the velocities for
existence of a causal Hamiltonian are derived without any assumption
about the functional form of the Hamiltonian.  When we substitute
Eqs. (15) for the velocities (given by Schr\"odinger Eqn.) into
Eqs. (22), we obtain $n(n-1)/2$ partial differential eqns. for the
$n(n-1)/2$ functions $W_{i\ell} (\vec x,t)$:
\begin{equation}
{\partial \over \partial x_i} \left({f_{kj} \over |\psi(\vec x,t)|^2}
{\partial W_{k\ell} \over \partial x_\ell}\right) - {\partial \over
\partial x_j} \left({f_{ki} \over |\psi(\vec x,t)|^2} {\partial
W_{k\ell} \over \partial x_\ell}\right) + F_{ij} = 0,
\end{equation}
where
\begin{equation}
F_{ij} \equiv {\partial f_{ij} \over \partial t} + {\partial \over
\partial x_i} (f_{kj} v_{k,B}) - {\partial \over \partial x_j} (f_{ki}
v_{k,B}).
\end{equation}
Except for very special wave functions (e.g. factorizable wave
functions), the $\hat p_i$ determined to fit quantum probability
distributions of $n+1$ CCS yield $f_{ij} \neq 0$ and $F_{ij} \neq 0$,
and Eqs. (24) do not have the trivial solution $W_{k\ell} = 0$.  In
contrast, in the dBB theory which reproduces \underbar{only} the
position probability density $(\hat p_i)_{\rm dBB} = \partial_i S$
which yields $f_{ij} = F_{ij} = 0$ and hence Eq. (24) is obeyed with
$W_{k\ell} = 0$.  Thus, for general wave functions, departure from dBB
velocities (i.e., $W_{k\ell} \neq 0$) is needed for existence of a
causal Hamiltonian when we insist on reproducing quantum probability
distributions of $n+1$ CCS $(n > 1)$ of observables. 
\bigskip

\noindent \underbar{Determination of Causal Hamiltonian}.  With
velocities so determined, $\hat H_c$ is found by integrating
Eqs. (20) without any assumption about the form of the Hamiltonian
$H_c$.  We shall show that a causal Hamiltonian exists without making
any claim of its uniqueness.  We now make the ansatz,
\begin{equation}
H_c (\vec x,\vec p,t) = \sum^n_{i=1} {(p_i - A_i(\vec x,t))^2 \over
2m_i} + V(\vec x,t).
\end{equation}
We find from Hamilton's eqns. (17),
$$
A_i (\vec x,t) = \hat p_i(\vec x,t) - m_i v_i
\eqno (26a)
$$
which yield the $A_i$; we then calculate $V(\vec x,t)$ from, 
$$
V(\vec x,t) = \hat H_c (\vec x,t) - \sum_i {(\hat p_i - A_i)^2 \over 2m_i}. 
\eqno (26b)
$$
This completes the determination of a causal Hamiltonian (26) which
contains $n+1$ quantum potentials.  The explicit forms of the
potentials for $n=1$ have been given in Ref. 9.  For $n=2$ they follow
from the velocity formula (15), and Eqs. (25) and (28).
\bigskip

\noindent\underbar{Explicit Formulae For Velocities For $n=2$}.  In 2
dimensions Eq. (24) simplifies to a first order partial differential
eqn. 
\begin{equation}
{\partial \over \partial x_2}(g_{12}) {\partial W_{12} \over \partial
x_1} - {\partial \over \partial x_1} (g_{12}) {\partial W_{12} \over
\partial x_2} - F_{12} = 0,
\end{equation}
where
$$
g_{12} (\vec x,t) = f_{12} (\vec x,t)/|\psi(\vec x,t)|^2.
\eqno (27a)
$$
Since no time derivatives occur, $t$ may be considered as a fixed
parameter and Eq. (27) solved by Lagrange's method by considering a
curve $x_1 = x_1(s)$, $x_2 = x_2(s)$ with
\[
ds = {dx_1 \over \partial g_{12}/\partial x_2} = - {dx_2 \over
\partial g_{12}/\partial x_1} = {dW_{12} \over F_{12}}.
\]
On this curve
\[
{dg_{12} \over ds} = 0, \ {dW_{12} \over ds} = F_{12}.
\]
Inserting back the fixed parameter $t$, we obtain the most general
solution, 
\begin{equation}
W_{12} (x_1,x_2,t) = h(g_{12},t) + \int^{x_1}_0 \left({F_{12} \over
\partial g_{12}/\partial x_2}\right) (x'_1,x'_2,t) dx'_1,
\end{equation}
where (i) the argument $x'_2$ along the path of integration equals
$x_2$ at $x'_1 = x_1$ and is determined for other values of $x'_1$ by
the condition $dg_{12} = 0$ along the curve, and (ii) $h$ is an
arbitrary function of $g_{12}$ and $t$.  The velocities given by (15)
will contain a degree of arbitrariness corresponding to the choice of
the function $h(g_{12},t)$.
\bigskip

\noindent 5. \underbar{Conclusions}.  The new causal quantum mechanics in
$n$ dim. configuration space has the following important
properties. (i) It reproduces quantum probability distributions of
$n+1$ CCS of observables with one positive definite phase space
density. (ii) It has a $c$-no. causal Hamiltonian which contains $n+1$
quantum potentials. (iii) It has velocities (and hence trajectories)
which (for general wave functions) differ from dBB velocities for $n >
1$ due to the insistence on reproducing quantum probabilities of $n+1$
CCS.  The velocities contain some arbitrariness (e.g. the function $h$
in 2 dim.) in spite of these constraints.  (iv) It has
position-momentum correlations in individual events (given by $\hat
{\vec p} (x,t)$) different from dBB theory.  Applications to quantum
chaos and possible experimental tests need further work.  We also hope
to compare numerically the trajectories implied by the present work
with those given by the de Broglie-Bohm theory in a future
communication. 
\bigskip

\noindent \underbar{Acknowledgements}.  We wish to thank members of
the theoretical physics department at TIFR, A. Martin at CERN,
A.P. Balachandran and R. Sorkin at Syracuse Univ., A. Fine and A. Garg
at Northwestern Univ., and I. Ohba, M. Namiki and H. Nakazato at
Waseda Univ. for discussions.  One of us (SMR) is grateful to
J.P. Vigier for encouragement during a symposium in his honour at York
Univ.  Part of the work was supported by U.S. DOE under contract
no. DE-FG02-85ER40231.

\newpage

\noindent{\bf References} \\

\begin{enumerate}
\item[{1.}] A.M. Gleason, {\it J. Math. \& Mech.} \underbar{6}, 885
(1957).
\item[{2.}] S. Kochen and E.P. Specker, {\it J. Math. \& Mech.}
\underbar{17}, 59 (1967); N.D. Mermin, Phys. Rev. Lett. \underbar{65},
3373 (1990). 
\item[{3.}] J.S. Bell, {\it Physics} \underbar{1}, 195 (1964).
\item[{4.}] A. Martin and S.M. Roy, {\it Phys. Lett.} \underbar{B350},
66 (1995).
\item[{5.}] L. de Broglie, ``Nonlinear Wave Mechanics, A Causal
Interpretation'', (Elsevier 1960); D. Bohm, {\it Phys. Rev.}
\underbar{85}, 166; 180 (1952); D. Bohm and J.P. vigier, {\it
Phys. Rev.} \underbar{96}, 208 (1954); D. Bohm, B.J. Hiley and
P.N. Kaloyerou, {\it Phys. Rep.} \underbar{144}, 349 (1987);
D. D\"urr, S. Goldstein and N. Zanghi, {\it Phys. Lett.}
\underbar{A172}, 6 (1992) \& {\it Found. of Phys.} \underbar{23}, 721
(1993); P.R. Holland, ``The Quantum Theory of Motion'' (Cambridge
Univ. Press, 1993); D. Bohm and B.J. Hiley, ``The Undivided
Universe'', (Routledge, London, 1993).
\item[{6.}] J.S. Bell, {\it Phys. Rep.} \underbar{137}, 49 (1986), and
in ``Speakable \& Unspeakable in Quantum Mechanics'' (Cambridge
Univ. Press, 1987), p. 173.
\item[{7.}] T. Takabayasi, {\it Prog. Theor. Phys.} \underbar{8}, 143 (1952).
\item[{8.}] S.M. Roy and V. Singh, `Deterministic Quantum Mechanics in
One Dimension', p. 434, Proceedings of International Conference on
Non-accelerator Particle Physics, 2-9 January, 1994, Bangalore (World
Scientific, 1995), R. Cowsik Ed. 
\item[{9.}] S.M. Roy and V. Singh {\it
Mod. Phys. Lett.} \underbar{A10}, 709 (1995).
\item[{10.}] E. Wigner, {\it Phys. Rev.} \underbar{40}, 749 (1932).
\item[{11.}] K. Husimi, {\it Proc. Phys. Math. Soc. Japan}
\underbar{22}, 264 (1940), R.F. O'Connell and E.P. Wigner, {\it
Phys. Lett.} \underbar{85A}, 121 (1981).
\item[{12.}] L. Cohen and Y.I. Zaparovanny, {\it J. Math. Phys. }
\underbar{21}, 794 (1980); L. Cohen, {\it J. Math. Phys.}
\underbar{25}, 2402 (1984).
\item[{13.}] E. Deotto and G.C. Ghirardi, Foundations of Physics,
\underbar{28}, 1 (1998); P.R. Holland, ibid. \underbar{28}, 881 (1998).
\end{enumerate}

\end{document}